\documentclass[a4paper,fleqn]{cas-sc}
\usepackage[numbers]{natbib}

\newcommand{\beq}{\begin{equation}}

\newcommand{\eeq}{\end{equation}}
\newcommand{\ee}[1] {\label{#1} \end{equation}}
\newcommand{\bea}{\begin{eqnarray}}

\newcommand{\eea}{\end{eqnarray}}
\newcommand{\barr}{\begin{array}}
\newcommand{\earr}{\end{array}}

\newcommand{\rf}     [1] {~\cite{#1}}
\newcommand{\refref} [1] {Ref.~\cite{#1}}

\newcommand{\reffig} [1] {Fig.~\ref{#1}}

\newcommand{\refFig} [1] {Figure~\ref{#1}}

\newcommand{\integers}{\ensuremath{\mathbb{Z}}}

\renewcommand{\det}{\mathrm{det}\,}

\newcommand{\ignore}[1]{}

\newcommand{\ssp}{\ensuremath{\phi}}             

\newcommand{\ExpaEig}{\ensuremath{\Lambda}}

\newcommand{\zeit}{\ensuremath{t}}  

\newcommand{\jMps}{\ensuremath{J}}   
\newcommand{\jMorb}{{\ensuremath{\mathcal{\jMps}}}}

 \newcommand{\lsts}{periodic states}

\newcommand{\pcell}{primitive cell}              

\newcommand{\jacobianOrb}{orbit Jacobian matrix}

\newcommand{\jacobianOp}{orbit Jacobian operator}
\newcommand{\JacobianOp}{Orbit Jacobian operator}

\newcommand{\Hilldet}{orbit Jacobian}

\newcommand{\speriod}[1]{{\ensuremath{L_{#1}}}}  

\newcommand{\steady}{steady state}
\newcommand{\Steady}{Steady state}

\newcommand{\dmn}{-dimensional}  

\iffalse
\ifboyscout 
  \newcommand{\PC}[2]{$\footnotemark\footnotetext{Predrag #1: #2}$}

  \newcommand{\DL}[2]
                     {\begin{quote}{\color{green}[#1 Domenico] \small #2}\end{quote}}
  
  \newcommand{\CPD}[2]{$\footnotemark\footnotetext{#1 Carl: #2}$}
  
  \newcommand{\CBlibrary}[1]
             {\href{https://ChaosBook.org/library/#1.pdf} { (click here)}}
\else 
  \newcommand{\PC}[2]{}{}
  
  \newcommand{\DL}[2]{}{}
  
  \newcommand{\CPD}[2]{}{}
  
  \newcommand{\CBlibrary}[1]{}
\fi  

\newcommand{\HREF}[2]{{\href{#1}{#2}}}

\newcommand{\toCLpdf}[2]{\HREF{https://arxiv.org/pdf/2503.22972v1\##1}{CL18\, #2}}
\newcommand{\toLCpdf}[2]   
    {\HREF{https://ChaosBook.org/overheads/spatiotemporal/LC21.pdf\##1}{LC21\, #2}}

\def\tsc#1{\csdef{#1}{\textsc{\lowercase{#1}}\xspace}}
\tsc{WGM}
\tsc{QE}
\tsc{EP}
\tsc{PMS}
\tsc{BEC}
\tsc{DE}

\begin{document}
\let\WriteBookmarks\relax
\def\floatpagepagefraction{1}
\def\textpagefraction{.001}
\shorttitle{Spatiotemporal stability of synchronized states}
\shortauthors{D. Lippolis}

\title [mode = title]
{Spatiotemporal stability of synchronized coupled map lattice states}



\author{Domenico Lippolis}[orcid=0000-0002-2817-0859] 
%
\ead{lippolis@tmu.ac.jp}
\ead[url]{https://cns.gatech.edu/~domenico/}


\affiliation{organization={Department of Physics, Tokyo Metropolitan University},
		  city={Hachioji}, 
                addressline={}, 
               citysep={},  
                postcode={192-0397},                
                country={Japan}}








\begin{abstract}
In the realm of spatiotemporal chaos, unstable periodic orbits play a 
major role in understanding the dynamics. 
Their stability changes and bifurcations in general are thus of central 
interest.    
Here, coupled map lattice discretizations of 
nonlinear partial differential equations, exhibiting a variety of 
behaviors depending on the coupling strength, are considered.   
In particular, the linear stability analysis of synchronized states is 
performed by evaluating the Bravais lattice {\Hilldet} in its 
reciprocal space first Brillouin zone,
with space and time treated on 
equal grounds. 
The eigenvalues of the {\jacobianOp}, computed as functions of the 
coupling strength, 
tell us 
about the stability of the periodic orbit under a perturbation of a 
certain time- and space frequency. Moreover, the stability under
aperiodic,
that is, incoherent perturbations, is revealed by 
integrating the sum of the stability exponents over all space-time 
frequencies.

\end{abstract}


\begin{highlights}
\item A spatiotemporal {\Hilldet} is computed for coupled map lattices in the first Bruillouin zone.
\item The linear stability of synchronized states under periodic and aperiodic perturbations is studied. 
\item Short periodic orbits exhibit a nontrivial dependence of the stability exponent on the lattice coupling.
\end{highlights}


\maketitle

\section{Introduction}
Consider a scalar field $\varphi(\vec x,t)$ described by
a reaction-diffusion PDE
\beq
\frac{\partial}{\partial t}\varphi=D\cdot\Delta\,\varphi + V'(\varphi)
\,,
\ee{GroBec06:sto2}
where the linear part is the {heat equation},
$D$ is a diffusion constant or a diffusion tensor,
and $V$ is a potential.
In one space dimension the Laplacian  is
$\Delta=\partial^2/\partial{x}^2$.
Ginzburg-Landau type equations and nonlinear Schr\"odinger equations
are examples of such systems.
Nonlinear PDEs of this kind may showcase a wide variety of dynamical behaviors, from integrability, to 
Turing instabilities, to full-fledged spatiotemporal chaos, and may serve as a model to describe phenomena
of virtually every energy scale.

Space-time discretizations of reaction-diffusion PDEs can make integration more practical, and the analysis
more insightful\rf{frisch86}. Out of the many realizations in the literature,  
Kaneko's `diffusive coupled map lattice' spacetime 
discretization of \eqref{GroBec06:sto2} is considered here\rf{Kaneko83,Kaneko84}. One makes the discretized 
coordinates an integer hypercubic lattice $\integers^{d}$ by scaling the 
lattice constants $\Delta{x}$, $\Delta{\zeit}$ to unit spacing, and by 
rescaling $\varphi(x,t)$ into a dimensionless field 
$\ssp_{n,\zeit}$.
The resulting discretized defining equation \eqref{GroBec06:sto2} is
\beq
\ssp_{n,\zeit+1} - \ssp_{n,\zeit}  = 
     \frac{a}{2}
     \left(\ssp_{n+1,\zeit} - 2\ssp_{n,\zeit} + \ssp_{n-1,\zeit}\right)
   + (1-a)(F(\ssp_{n,\zeit}) - \ssp_{n,\zeit})
\,.
\ee{Dettmann02:e:maps}
Here $(n,\zeit)\in\integers^2$ are integer square lattice spacetime 
positions, 
$a$ is the dimensionless diffusion constant $a=2D\,\Delta\zeit/(\Delta{x})^2$, 
the Laplacian couples neighboring lattice site fields with strength $a$, 
and $F$ is forward-in-time map at each site ${n\zeit}$, related to the 
potential in \eqref{GroBec06:sto2} by 
\beq 
V'(\ssp_{n,\zeit})=  (1-a)(F(\ssp_{n,\zeit}) - \ssp_{n,\zeit})
\,.
\ee{GroBec06:map}
 
As the correspondence between Eqs.~(\ref{GroBec06:sto2}) and~(\ref{Dettmann02:e:maps}) shows, coupled map lattices lend themselves to 
approximating
less tractable continuous systems, such as those describing 
turbulence\rf{bohr98tur,BarkleyCML11,ChanTuckBar17}. They have been studied 
theoretically\rf{BunSin88,MacKayCML00,ChaFer05}, phenomenologically as paradigms of pattern formation\rf{KilLofSun13}, as well as experimentally\rf{CMLChimera12}. 
Farther-reaching applications include neural networks\rf{Nozawa92}, cryptography\rf{Tenny03}, and most recently, reservoir computing\rf{LanResComp26}.   

In his chaotic quantization of field theories\rf{Beck95}, C. Beck introduces the coupled map lattice~(\ref{Dettmann02:e:maps})
with the map $F(\phi)$ in Eq.~(\ref{GroBec06:map}) 
given by the Chebyshev polynomials 
\beq
T_N(\phi) = \cos\left(N\arccos\phi)\right)
\,, \hspace{1cm} |\phi|\leq1 \,,
\label{ChebPol} 
\eeq
in order to replace Gaussian noise by deterministic chaos in the Parisi-Wu stochastic quantization\rf{ParWu81}.
Chebyshev maps, as constituents of the many-body chains~(\ref{Dettmann02:e:maps}) evolving in discrete time, are smoothly conjugated to
Bernoulli shifts and thus strongly chaotic. In spite of that, stable 
synchronized states at small couplings have been 
discovered~(\citet{Dettmann02}) for this model. 

The emergence of coherent structures\rf{BunLamLi90} and the study of instability fields\rf{IPRT90,LePoTo96,PaSzLoRo09,JiPoTo13,DubShu25} in coupled map lattices
have been issues of great relevance for decades, due to the rich dynamical landscape offered by these models, as well as the effectiveness of such tools as the properly defined Lyapunov exponents and Lyapunov vectors. 

Recently, the study of the stability of spatiotemporal recurrent patterns has found application in chaotic lattice field theories\rf{LC21,CL18}, whose partition functions are written as sums over periodic orbits in spacetime geometries. The weights of the periodic orbits depend on their instabilities, evaluated by linearizing the dynamics on a lattice, where spatial and temporal directions are treated on equal footing, and evaluated on the reciprocal lattice, by means of Bloch's theorem.  

The operation of Fourier-transforming to a wavenumber-\-frequency space (in both real space and time) emphasizes an aspect of dynamics, which is at the center of the present study:
linear stability analysis as we know it deals with the response of a \textit{periodic} orbit to \textit{periodic} perturbations, best characterized in reciprocal space by their frequencies.
Yet, perturbations are in general
aperiodic,
or incoherent, and may be obtained from superpositions of all frequencies, for a thorough stability analysis.    

Here, the stability properties are investigated in a special family of periodic orbits, the synchronized states\rf{PecCar90,WeakSynchCML98} of the evolving chain~(\ref{Dettmann02:e:maps}).     
In particular, the goal of the analysis is to describe the behavior of the stability exponents of the synchronized states as a function of 
the lattice coupling stength,
as well as to demonstrate that short periodic orbits become stable\rf{Dettmann02} under coherent and incoherent perturbations, for large enough couplings. 

The article is organized as follows: in section~\ref{TvsST}, the differences are explained between the time-forward and the spatiotemporal approaches to linear stability analysis;
in section~\ref{BlochJac}, 
the spatiotemporal {\jacobianOp} in reciprocal space is introduced; the spatiotemporal linear stability analysis is then performed on the synchronized states
of the coupled Chebyshev map lattices in section~\ref{SynchStatesStab}, and in particular on steady states (section~\ref{SteadStat}), as well as on period-2 orbits (section~\ref{p2UPO}). Conclusions are drawn in section~\ref{end}.   

\section{Forward-in-time- versus spatiotemporal {\jacobianOp}}
\label{TvsST} 
Linear stability is typically evaluated through the familiar forward-in-time Jacobian, obtained as follows.
First, rewrite the discretized defining equation~(\ref{Dettmann02:e:maps}) 
as a
forward-in-time
map:
\beq 
 \ssp_{n,\zeit+1} = 
    (1-a)T_N(\ssp_{n,\zeit}) 
    + \frac{a}{2}(\ssp_{n-1,\zeit} + \ssp_{n+1,\zeit}) 
\,. 
\ee{ChebCML}
Here, the $N$th Chebyshev polynomial $T_N(\ssp)$ defined in~(\ref{ChebPol}) replaces the general $F(\ssp)$
in Eq.~(\ref{Dettmann02:e:maps}). 
The forward-in-time evolution of perturbations is given by
\beq
\delta\Phi_{t+1} = J_t \delta\Phi_t
\,,
\hspace{0.5cm}
[J_t]_{mn} = \frac{\partial \phi_{m,t+1}}{\partial \phi_{n,t}} 
\,,
\ee{PertEvol}  
where $\Phi_t=(\phi_{0,t},...,\phi_{L-1,t})^T$, 
and the $L\times L$ temporal Jacobian is
\beq
 J_t = \left( 
\begin{array}{cccccc}
 (1-a)T_N'(\phi_0) & \frac{a}{2} & 0 & 0 & ... & \frac{a}{2} \\
 \frac{a}{2} & (1-a)T_N'(\phi_1) & \frac{a}{2} & 0 & ... & 0 \\
 0 &  \frac{a}{2} & (1-a)T_N'(\phi_2) &  \frac{a}{2} & ... & 0 \\
... & ... & ... & ... & ... & ...     \\
 \frac{a}{2} & 0 & 0 & ...  &  \frac{a}{2} & (1-a)T_N'(\phi_{L-1}) 
\end{array} \right)
\,.
\eeq
In this temporal picture, the full stability matrix of an orbit $\{\Phi_0,\Phi_1,... ,\Phi_{\tau-1}\}$ is the product of the 
$J_t$'s computed at each point,
\beq
J = J_{\tau-1}J_{\tau-2}\cdot\cdot\cdot J_0
\,.
\ee{TempJac}
 The stability of $J$ then depends on its eigenvalues.
 
 On the other hand, 
the {\jacobianOp}\rf{Bount81} describes the linearized evolution on the full spatiotemporal  lattice, and it is obtained 
 by differentiating Eq.~(\ref{Dettmann02:e:maps}):   
 \beq
{\jMorb}_{n'\zeit';n\zeit} = 
  (\delta_{\zeit',\zeit+1} - \delta_{\zeit',\zeit})\delta_{n',n} 
  - \frac{a}{2}(\delta_{n',n-1} -2\delta_{n',n} + \delta_{n',n+1})\delta_{\zeit',\zeit}
  - V''(\ssp_{n,\zeit})\delta_{n',n}\delta_{\zeit',\zeit} 
\,,
\label{2dOrbJac}
\eeq
where $V''(\phi)$ is given by Eq.~(\ref{GroBec06:map}) with $F=T_N(\phi)$. If we consider a spatiotemporal cell of size $[L\times \tau]$, 
the {\jacobianOrb} $\jMorb$ is a $L\tau\times L\tau$ matrix, much larger in general than the temporal Jacobian $J$. 
If the orbit $\{\Phi_0,\Phi_1,... ,\Phi_{\tau-1}\}$ is periodic, 
the determinants of 
the temporal and {\jacobianOp}s are related by Hill's formula\rf{LC21}
\beq
|\det{\jMorb}| = |\det{(1-J)}|
\,.
\ee{Hills}
For the sake of clarity, let us compare these two notions of stability in the simple example of a single-site spatiotemporal lattice.

\subsection{One spatial lattice site Jacobian}
Consider a one\dmn\ lattice, infinitely extended in time, 
but with only one site in the spatial dimension. In this case, the 
{\jacobianOrb}~(\ref{2dOrbJac}) reduces to 
\beq {\jMorb}_{\zeit';\zeit} = 
    \delta_{\zeit',\zeit+1} - \delta_{\zeit',\zeit} 
    - V''(\ssp_{\zeit})\,\delta_{\zeit',\zeit} \,. 
\ee{1dOrbJac}
For a periodic orbit of temporal period $\tau$, Eq.~(\ref{1dOrbJac}) 
reads, in matrix form, 
\beq
 {\jMorb} = \left( 
\begin{array}{cccccc}
 -1-V_0'' & 1 & 0 & 0 & ... & 0 \\
 0 & -1-V_1'' & 1 & 0 & ... & 0 \\
 0 & 0 & -1-V_2'' & 1 & ... & 0 \\
... & ... & ... & ... & ... & ...     \\
1 & 0 & 0 & 0  & ...& -1-V_{\tau-1}'' 
\end{array} \right)
\,.
\eeq
 The determinant of ${\jMorb}$ can be expanded with respect to the first column, and so written as a sum of two circulants:
 \beq
 \det{\jMorb} = (-1-V_0')K_{\tau-1} + (-1)^{\tau-1} K_{1,\tau-1} 
 \,,
 \label{1ddet}
 \eeq
 with
 \beq
 K_{\tau-1} = \left|
 \begin{array}{cccccc}
 -1-V_1'' & 1 & 0 & 0 & ... & 0 \\
 0 & -1-V_2'' & 1 & 0 & ... & 0 \\
 0 & 0 & -1-V_3'' & 1 & ... & 0 \\
... & ... & ... & ... & ... & ...     \\
0 & 0 & 0 & 0  & ...& -1-V_{\tau-1}'' 
\end{array} \right|
\,,
\eeq
and
\beq
 K_{1,\tau-1} = \left|
 \begin{array}{cccccc}
 1 & 0 & 0 & 0 & ... & 0 \\
 -1-V_1'' & 1 & 0 & ... & 0 \\
 0 & -1-V_2'' & 1 & ... & 0 \\
... & ... & ... & ... & ... & ...     \\
0 & 0 & 0 & ... & -1-V_{\tau-2}'' & 1 
\end{array} \right|
\,.
\eeq
The previous are both determinants of triangular matrices, and thus Eq.~(\ref{1ddet}) is finally
\beq
 \det{\jMorb} = \prod_{t=0}^{\tau-1}(-1-V_t'') + (-1)^{\tau-1} = (-1)^{\tau-1} + (-1)^\tau\prod_{t=0}^{\tau-1}[a+(1-a)T_N'(\phi_t)] 
 \,.
 \label{1ddet_fin}
 \eeq
In order to compare Eq.~(\ref{1ddet_fin}) with the result $|\det{(1-J)}|$ from the temporal calculation, recall the map~(\ref{ChebCML}) for a one-site lattice
\beq
\phi_{t+1} = (1-a)T_N(\phi_{t}) + a\phi_{t}
\,.
\label{ChebCML1d}
\eeq
The temporal Jacobian is one-dimensional and simply given by the product of the derivatives of the above expression, resulting in the determinant 
\beq
\det{(1-J)} = 1 -  \prod_{t=0}^{\tau-1}[a+(1-a)T_N'(\phi_t)]
\,,
\eeq 
which is the same as~(\ref{1ddet_fin}) in absolute value.

\section{{\JacobianOp} in reciprocal space}
\label{BlochJac}
The Jacobian determinant is associated to the stability of a periodic orbit of period $\tau$ to a perturbation of the same period, or of  
 a multiple of it (repeat). Equivalently, one can switch to a reciprocal lattice where periods are replaced by frequencies, just like it is 
 usually done in condensed matter physics with Bravais lattices. The advantage of using reciprocal space is at least twofold: first, this 
 representation accounts for cyclic translations\rf{CL18} of the same orbit, and therefore automatically operates a partial symmetry reduction
 of the periodic orbits on the lattice; secondly, as it will be explained below for synchronized states, the stability exponents of each orbit 
 can be summed over perturbations of all frequencies, obtaining the stability under incoherent, that is aperiodic, perturbations.
 
 The first goal of the stability analysis in the reciprocal lattice is to obtain the eigenvalues of the Jacobian for a given 
 perturbation frequency. In the temporal representation, this is most easily achieved by searching for complex eigenvalues of the cyclic matrix $J$ [Eq.~(\ref{TempJac})]
 of the form\rf{Dettmann02}
 \beq
 \ExpaEig_k = \sum_{n=0}^{L-1} b_n e^{ink}
 \,,
 \eeq
 with $k=2\pi n/L$,
in a space of eigenvectors of the form $v_k=(1, e^{ik}, e^{2ik}, ..., e^{-ik})^\top$. Applying $J v_k = \ExpaEig_k v_k$ yields
\beq
\ExpaEig_k  = \prod_{t=0}^{\tau-1} \left[(1-a)T_N'(\phi_t) + a\cos k\right]
\label{sstempeig}
\,.
\eeq
  
On the other hand,  in the spatiotemporal picture, the eigenvalues and the determinant of the {\jacobianOrb} 
in reciprocal space are obtained by first taking the discrete Fourier transform  $\tilde{{\jMorb}}_{m,m'}$ [with $m=(m_1,m_2)]$
of the matrix  element $\jMorb_{n't';nt}$ [Eq.~(\ref{2dOrbJac})]: 
\begin{flalign}
\nonumber
&\tilde{{\jMorb}}_{m,m'} = 
\sum_{n' n t' t} e^{i(k_1t - k_1't')+i(k_2n - k_2'n')}(\delta_{t',t+1} - \delta_{t',t})\delta_{n',n} -\sum_{n' n t' t} e^{i(k_1t - k_1't')+i(k_2n - k_2'n')}V''(\phi_{nt})\delta_{n',n}\delta_{t',t}  \\
\nonumber
&-\frac{a}{2}\sum_{n' n t' t} e^{i(k_1t - k_1't)'+i(k_2n - k_2'n')}(\delta_{n',n-1} -2\delta_{n',n} + \delta_{n',n+1})\delta_{t',t} =  \\ \nonumber =
&\sum_{n,t} e^{i(k_1-k_1')t + i(k_2-k_2')n}(e^{-ik_1'}-1) -\sum_{n,t} e^{i(k_1-k_1')t + i(k_2-k_2')n}V''(\phi_{n,t}) 
\\ &-\frac{a}{2}\sum_{n,t} e^{i(k_1-k_1')t + i(k_2-k_2')n}(e^{ik_2'} -2 + e^{-ik_2'})   \\   
&=(e^{-ik_1'}-1)\delta_{m',m} -\tilde{V}''(\phi) -a(\cos k_2' -1)\delta_{m',m} \,,
\label{Jmn}
\end{flalign}
where the term 
$\tilde{V}''(\phi)=\sum_{n,t} e^{i(k_1-k_1')t + i(k_2-k_2')n}V''(\phi_{n,t})$ 
produces the non-diagonal entries in the reciprocal Jacobian. Here 
$k_1=\frac{2m_1\pi}{\tau}$, $k_2=\frac{2m_2\pi}{L}$, $m=(m_1,m_2)$, 
with $m_1=0,...,\tau-1$, and $m_2=0,...,L-1$, for 
a rectangular \pcell\ $[{\speriod{}}\times{\tau}]$ of spatial 
period $\speriod{}$ and temporal period $\tau$. Then one computes the spectrum of 
the {\jacobianOp} $\tilde{{\jMorb}}$,
which is now shown in the special case of synchronized states, and compared to the temporal result~(\ref{sstempeig}).

\section{Spatiotemporal stability of synchronized states}
\label{SynchStatesStab}
Synchronized states are solutions in which
\(
\ssp_{n,\zeit} = \ssp_{\zeit}
\)
is independent of the spatial position $n$, so are given by a one-dimensional map.
They are the simplest bridge from 1\dmn\ systems to their
2\dmn\ transverse modes, with continuous $k$ stability.
In chaotic field theory, the focus is on \emph{unstable} {\lsts}. 
Among those, short temporal period solutions in particular are closely related
to the unstable ones. 

Regarding our model~(\ref{Dettmann02:e:maps}), it was established in 
\refref{Dettmann02} 
 that what makes synchronized states  still quite different from one\dmn\ temporal
lattice systems is their spacetime stability.
It was found earlier\rf{AGGN91,GadAmr93} for generic coupled map lattices, that the linear stability properties 
of the single-site periodic orbits carry over to
the corresponding synchronized states (therein called homogenous solutions)
only for temporal period $\tau=1$, that is what we call fixed points or steady states.
For cycles of longer periods, the stability of  the synchronized states may in general differ     
from that of the 1\dmn\ system.

The linear stability properties of the full coupled map lattice are
slightly different from that of the synchronized map, so that although
stability occurs in the region of synchronized superstable orbits, the
superstable orbit may not be linearly stable in the extended system.
The difference between stability in the synchronized map and the extended
system is that the full coupled map lattice has an infinite number of
degrees of freedom, leading to a richer spectrum of possible
instabilities.
Linear stability follows if all multipliers are contracting, for all
\speriod{} discrete  eigenmodes. 

In the next sections, the linear stability of time period-1 (steady states) and period-2 synchronized states is investigated of 
the coupled map~(\ref{Dettmann02:e:maps}), to periodic perturbations of all frequencies, or to aperiodic   (incoherent) 
perturbations.

\subsection{{\Steady}s}
\label{SteadStat}
Let us begin from the \steady, where the field has the same value across the lattice, and the term of self-interaction potential in the Jacobian~(\ref{2dOrbJac}) is constant: 
\beq
V''(\phi_{n,t}) = V''(\phi_t) = (1-a)(T_N'(\phi_t)-1)
\,.
\eeq 
The synchronized steady state solves fixed-point condition for the map~(\ref{ChebCML}) with $\ssp_{n,\zeit} = \ssp_{\zeit}$:  
\beq 
 \ssp_{\zeit} = 
    (1-a)T_N(\ssp_{\zeit}) 
    + \frac{a}{2}(\ssp_{\zeit} + \ssp_{\zeit}) 
\,. 
\ee{ChebCMLSS} 
For the Chebyshev $T_2(\phi)=2\phi^2-1$ model, there are two fixed points, that is $\phi_0=-1/2$ and $\phi_1=1$.
Let us denote either fixed point with $\phi_p$ in the following analysis.
At each steady state, the reciprocal {\jacobianOrb}~(\ref{Jmn}) is diagonal, since
\beq
\sum_{n,t} e^{i(k_1-k_1')t + i(k_2-k_2')n}V''(\phi_{n,t}) = V''(\phi_p)\sum_{n,t} e^{i(k_1-k_1')t + i(k_2-k_2')n} =   (1-a)(T_N'(\phi_p)-1)\delta_{m',m}
\,,
\eeq
and, overall, 
\beq
\tilde{{\jMorb}}_{m,m'} =  \left[e^{-ik_1}-(1-a)T_N'(\phi_p)-a\cos k_2 \right]\delta_{m',m}
\label{sseig}
\,.
\eeq
Eq.~(\ref{sseig}) is also the expression for the eigenvalues of the 
{\jacobianOrb}, that can now be compared with those\rf{Dettmann02} of 
the temporal Jacobian $J$ in Eq.~(\ref{sstempeig})  with $\tau=1$: 
\beq
\ExpaEig_k  =  (1-a)T_N'(\phi_p) + a\cos k
\,.
\label{sstempeig2}
\eeq
 If, in the spatiotemporal picture, we choose the simplest primitive cell $[L\times1]$ for this steady state, we have $\tau=1$, $m_1=0$, and the
 phase  $e^{-ik_1}=1$.  
 Then, identifying the $k_2$ of Eq.~(\ref{sseig}) with the $k$ of Eq.~(\ref{sstempeig2}), $|\tilde{{\jMorb}}_{m,m}|$ is the same as $|1- \ExpaEig_k|$
 from the temporal Jacobian.
 Equation~(\ref{sseig}) reveals the stability of the time period-1 synchronized state $\Phi_p$ under perturbations of time frequency $k_1$ and 
 space frequency $k_2$: unstable if $|\Lambda_{k_2,k_1}|=|\tilde{{\jMorb}}_{m,m}|>1$, stable if  $|\Lambda_{k_2,k_1}|<1$, marginally stable if
 $|\Lambda_{k_2,k_1}|=1$. Analogously, the expression~(\ref{sstempeig2}) for the single, $k$-dependent
eigenvalue, should be greater than unity in absolute value. That is equivalent to requiring that
\beq
T_N'(\phi_p) > \frac{1-a\cos k}{1-a}
\,, \hspace{1cm} \mathrm{or} \hspace{1cm}
T_N'(\phi_p) < -\frac{1+a\cos k}{1-a}
\,,
\label{Dett_stab}
\eeq  
depending on whether $T_N'(\phi_p)>0$ or $T_N'(\phi_p)<0$.
\begin{figure}[tbh!]
 \centerline{
(a)\scalebox{.55}{\includegraphics{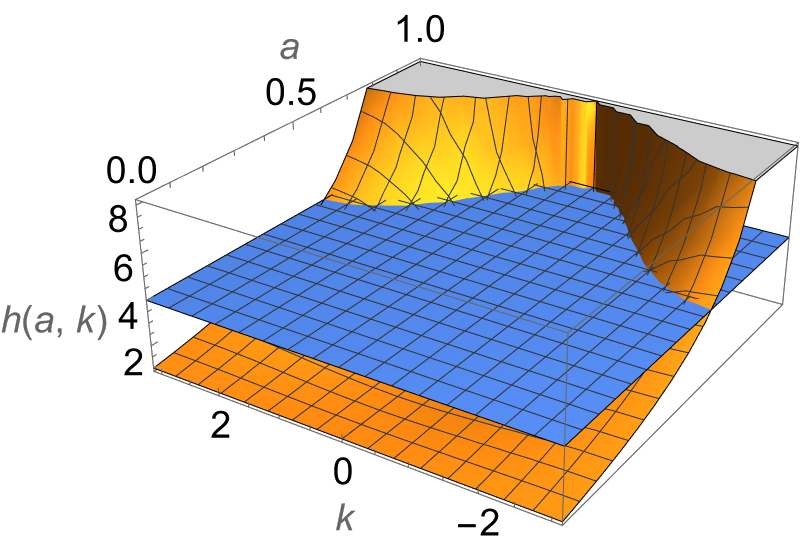}}
(b)\scalebox{.55}{\includegraphics{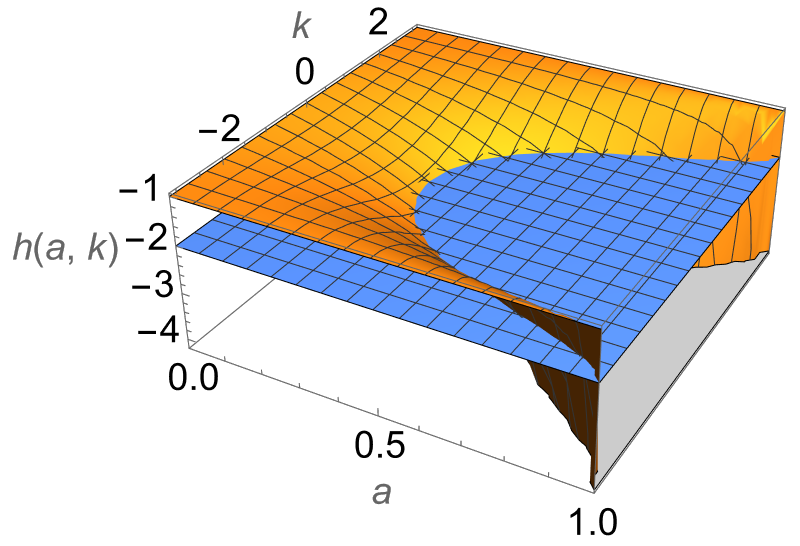}}}
\caption{The range of $a$ and $k$ that makes $\Lambda_k$ unstable is the region where the plane $z=T_N'(\phi_p)$ (in blue) is: (a) higher than the surface $h(a,k)= \frac{1-a\cos k}{1-a}$ (in orange) in the $T_2$ model with steady state $\phi_1=1$; (b) lower than the surface $h(a,k)= -\frac{1+a\cos k}{1-a}$ (in orange), $T_2$ model with steady state $\phi_0=-1/2$.}
\label{fakTN}
\end{figure}
\reffig{fakTN} tells us that, for small coupling $a$, all eigenvalues are unstable, and thus the steady state $\phi_p$ is unstable to perturbations of all space-frequencies.
However, for large enough $a$ the inequalities~(\ref{Dett_stab}) no longer hold for all $k$'s, and so stable eigenvalues appear, meaning that the steady state is
stable to perturbations of a certain frequency range $[-k^*,k^*]$.   
 More and more eigenvalues become stable as $a$ increases. 
 In the limit $a\rightarrow1$, the right-hand side of the inequalities~(\ref{Dett_stab}) diverges, that is, in order to achieve
mode instability in the limit of strongest coupling, it is required an infinite local stability multiplier.

\begin{figure}
 \centerline{
(a)\scalebox{.55}{\includegraphics{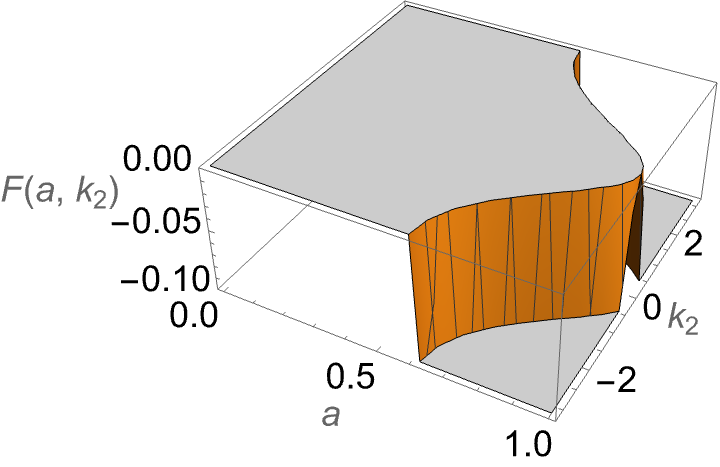}}
(b)\scalebox{.55}{\includegraphics{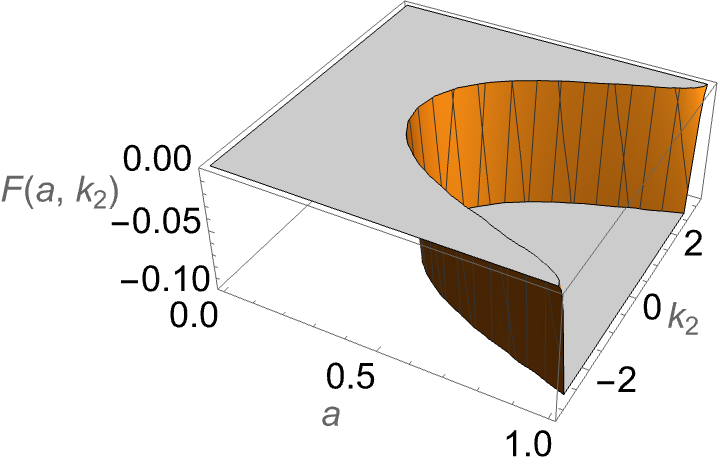}}}
\caption{The range of $a$ and $k_2$ that make $|f(a,k_2)|<1$  is the interval where the function $F(a,k_2)=|f(a,k_2)|-1$ is negative. (a) $T_2$ model and $\phi_1=1$. (b) $T_2$ model and $\phi_0=-1/2$.}
\label{fak2TN}
\end{figure}

 The stability of the synchronized steady state under incoherent (aperiodic) perturbations is best determined by 
 considering the spatiotemporal {\Hilldet}, and, in particular, by computing $\ln \det \jMorb$.
Equivalently, one can average the log of Eq.~(\ref{sseig}) over all frequencies\rf{CL18}, as 
 \beq
 \lambda = \frac{1}{4\pi^2} \sum_{k_1,k_2} \Delta k_1 \Delta k_2 \ln \Lambda_{k_2,k_1}
 \,.
 \label{BrStbExp}
 \eeq
 Recalling that $k_1=\frac{2m_1\pi}{\tau}$, $k_2=\frac{2m_2\pi}{L}$, the limit of the space and time periods
 of the primitive cell $L\rightarrow\infty$, $\tau\rightarrow\infty$ should be taken, with $m_1, m_2$ running
 from zero to infinity.
 The reason is that perturbations of a state can have the periodicity of the state itself, a different periodicity
 (or a different frequency in the Bravais lattice), or no periodicity at all~\cite{Pikovsky89}, that is a superposition
 of incoherent frequencies.
 In the summation~(\ref{BrStbExp}), $\Delta k_1=\frac{2\pi}{T}$ and $\Delta k_2 = \frac{2\pi}{L}$. 
 In the limit, the wavenumbers $k$ are continuous, and conventionally restricted to the first Brillouin zone
 $\mathbb{B} = \{k_1 \in (-\pi/T, \pi/T]\} \cup \{k_2 \in (-\pi/L, \pi/L]\}$, since the Bravais lattice is periodic and the $k-$vectors
 exceeding $\mathbb{B}$ do not carry any more information than their counterparts.        
 Then, the Bravais stability exponent~(\ref{BrStbExp}) is given as the
 integral 
 \beq
 \lambda =   \frac{1}{4\pi^2} \int_{-\pi}^\pi dk_2 \int_{-\pi}^\pi dk_1\ln \Lambda_{k_1,k_2}
 \,.
 \label{StbExp}
 \eeq  
 In the specific case of the synchronized steady state of the Chebyshev lattice, the previous reads
 \beq
 \lambda = \frac{1}{4\pi^2} \int_{-\pi}^\pi dk_2 \int_{-\pi}^\pi dk_1\ln [e^{-ik_1}-(1-a)T_N'(\phi_p)-a\cos k_2]
 \,.
 \eeq 
 The integral in $dk_1$ is of the form
 \beq
 I(a) = \int_{-\pi}^\pi dk_1\ln [e^{-ik_1} + f(a,k_2)]
 \,,
 \eeq
 with $f(a,k_2)=-(1-a)T_N'(\phi_p)-a\cos k_2$.
 This can evaluated by first differentiating with respect to $a$: 
 \beq
 \frac{dI(a)}{da} = \int_{-\pi}^\pi dk_1 \frac{f'(a,k_2)}{e^{-ik_1} + f(a,k_2)} = \left\{ 
 \begin{array}{cc} 
 \frac{2\pi f'(a,k_2)}{f(a,k_2)}\,, & |f(a,k_2)|>1 \\
 0\,, &    |f(a,k_2)|<1
 \end{array} \right.
 \,,
 \label{k1int}
 \eeq
as a result of the residue theorem: 
 \beq
  \frac{dI(a)}{da} =\oint_ {\Gamma}  \frac{dz}{iz} \frac{f'(a,k_2)}{z + f(a,k_2)}
  \,, \hspace{1cm}
  \vcenter{\includegraphics[width=0.2\linewidth]{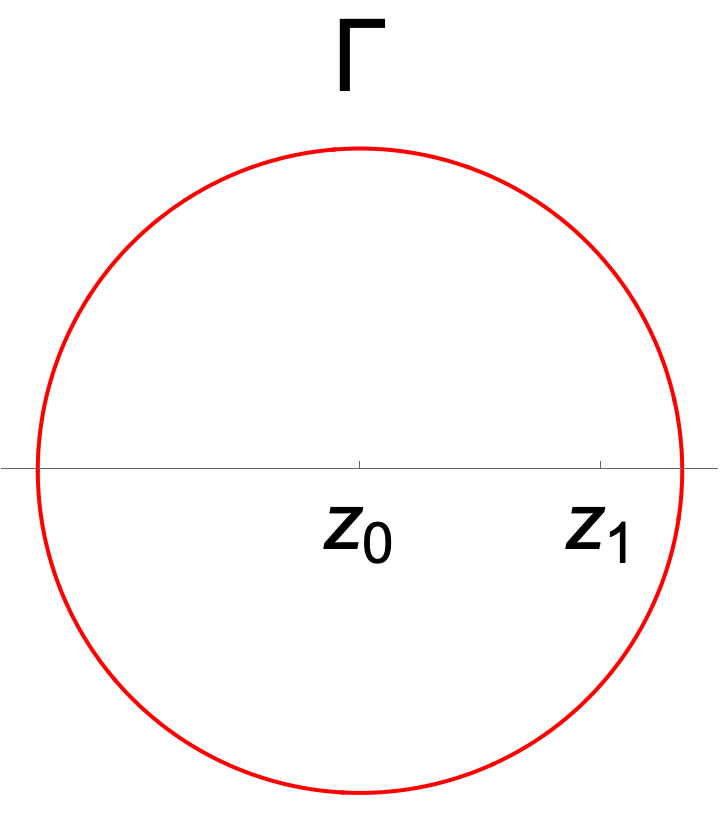}}
\label{dIa} 
\eeq
There are two poles at $z_0=0$ and $z_1=-f(a,k_2)$, and the latter is not enclosed by the contour $\Gamma$ if $ |f(a,k_2)|>1$, equivalent to
 \beq
 a < \mathrm{min}\left[ \frac{T_N'(\phi_p) -1}{T_N'(\phi_p) -\cos k_2}\,,
\frac{T_N'(\phi_p) +1}{T_N'(\phi_p) -\cos k_2} \right]
 \,,
 \label{k1intcond}
 \eeq
which depends on the sign of $T_N'(\phi_p)$.
If the previous inequality is verified, the pole $z_1$ lies outside $\Gamma$, and the integral is
\beq
  \frac{dI(a)}{da} = 2\pi i\, \mathrm{Res}(z_0) =  \frac{2\pi f'(a,k_2)}{f(a,k_2)}
 \,.
 \eeq
If, instead, $|f(a,k_2)|<1$, both poles contribute equal and opposite terms  to the contour integral:
 \beq
  \frac{dI(a)}{da} = 2\pi i\, \left[\mathrm{Res}(z_0) + \mathrm{Res}(z_1)\right] =  2\pi \left(\frac{f'(a,k_2)}{f(a,k_2)} -\frac{f'(a,k_2)}{f(a,k_2)}\right) = 0
 \,.
 \eeq
 In particular, the $k_2=\pi$ that maximizes the denominator of the above expression yields an upper bound for $a$, such that~(\ref{k1int}) is nonzero. Beyond that, the integral does not vanish in a nontrivial range of $k_2$, smaller than $[-\pi,\pi]$ (\reffig{fak2TN}).
Condition~(\ref{k1intcond}) is the same as~(\ref{Dett_stab}) for the stability of the eigenvalues of the Jacobian vs. space frequency. Here it is retrieved independently, and expressed as an inequality for  the lattice coupling strength $a$ (\textit{cf.} \reffig{fakTN}). 

\begin{figure}[tbh!]
 \centerline{
(a)\scalebox{.55}{\includegraphics{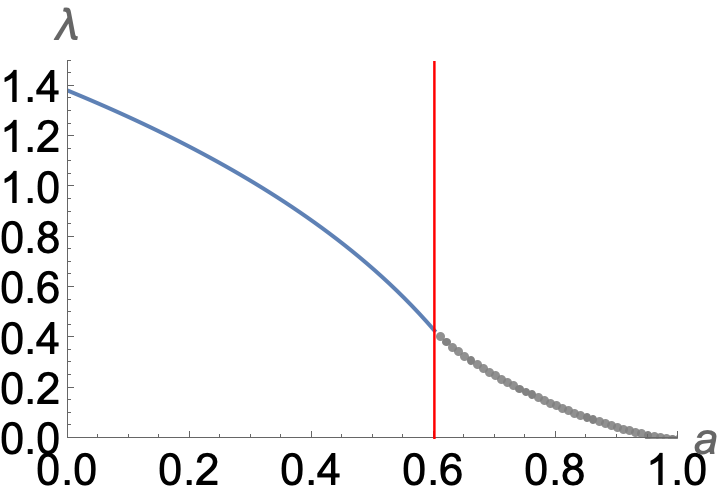}}
(b)\scalebox{.55}{\includegraphics{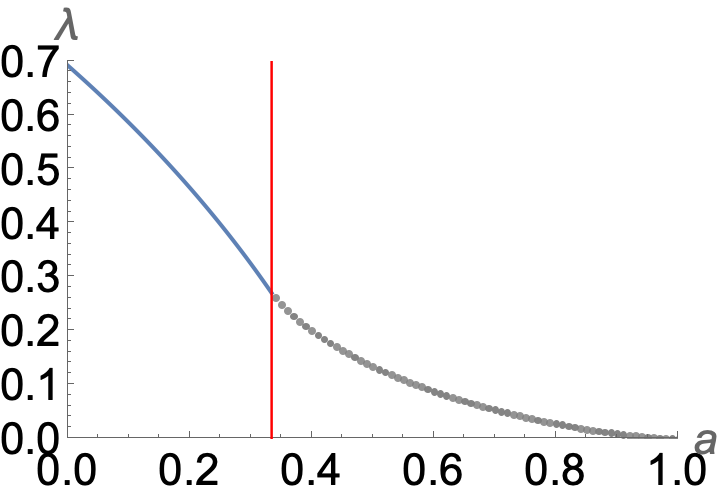}}}
\caption{The stability exponent versus strength of coupling $a$ (solid blue line). Beyond the red vertical line the condition $|f(a,k_2)|>1$ no longer holds and thus the stability exponent is evaluated 
numerically (gray dots). (a) $T_2$ model for the steady state $\phi_1=1$  (b) $T_2$ model for the steady state $\phi_0=-1/2$.}
\label{lamwholeT2}
\end{figure}

 Let us now continue to evaluate the stability exponent~(\ref{StbExp}). Suppose that~(\ref{k1int}) is nonzero, and so
 \beq
 I(a) =  2\pi \ln |f(a,k_2)| + c
 \,.
 \eeq
 Because $I(0)=\ln|T'_N(\phi_p)| + c$  and that should just be the stability exponent of the one-dimensional fixed point, it must be $c=0$.
 We are left with 
 \beq
 \lambda =  \frac{1}{2\pi} \int_{-\pi}^\pi dk_2 \ln|(1-a)T_N'(\phi_p)+a\cos k_2|
 \,.
 \label{k2int}
 \eeq
 The integral is computed in Appendix~\ref{AppA}, and the result is
 \beq
 \lambda = \ln \frac{(1-a)|T_N'(\phi_p)| + \sqrt{(1-a)^2T_N'^2(\phi_p) - a^2}}{2}
 \,,
\label{k2res}
 \eeq
 with the condition 
 \beq
 a<\frac{|T_N'(\phi_p)|}{1+|T_N'(\phi_p)|} 
\,.
\label{SScompcond}
\eeq
 When $a=0$  this again reduces to the one-dimensional stability exponent $\ln|T'_N(\phi_p)|$.
Let us now compare the ranges of validity~(\ref{k1intcond}) and~(\ref{SScompcond}) for the above result. In the models considered here, $T_2'(\phi)$ is such that 
\beq
 \frac{T_2'(\phi_1) -1}{T_2'(\phi_1) -\cos k_2}\,<\frac{T_2'(\phi_1)}{1+T_2'(\phi_1)}\,, \hspace{0.5cm} \mathrm{and} \hspace{0.5cm}
\frac{T_2'(\phi_0) +1}{T_2'(\phi_0) -\cos k_2} < \frac{T_2'(\phi_0)}{1-T_2'(\phi_0)}
\,,
\eeq
so the Bravais stability exponent is equal to~(\ref{k2res}) when $a$ is small enough to verify the inequality~~(\ref{k1intcond}). 
For larger couplings, the integral~(\ref{k2int}) has limits of integration $\pm k_2^*$, where
 \beq
 k_2^* = \arccos \frac{1-(1-a)T_N'(\phi_p)}{a}
 \,.
\label{k2st}
 \eeq
 A primitive function does not seem to exist~(discussion in Appendix~\ref{AppA}) for the integral in $k_2$, so that the stability exponent
  \beq
 \lambda =  \frac{1}{2\pi} \int_{-k_2^*}^{k_2^*} dk_2 \ln|(1-a)T_N'(\phi_p)+a\cos k_2|
 \label{k2intk2}
 \eeq
 is computed numerically, with the results portrayed in \refFig{lamwholeT2}. 

One can conclude that the steady states are always unstable under aperiodic perturbations, but their instability $\lambda$ 
decreases with the lattice coupling.
 When the condition~(\ref{k1intcond}) [or, equivalently, (\ref{Dett_stab}), red vertical line in the figure] for the single eigenvalues $\Lambda_{k_1,k_2}$ is no longer verified, some eigenvalues become stable, depending on the space frequency $k_2$. 
Then, the Bravais stability exponent $\lambda$ drops further, with the steady states tending towards stability ($\lambda\rightarrow0$) as $a\rightarrow1$.   
The observed behavior of the Bravais stability exponent reflects our qualitative understanding of the spatiotemporal lattice dynamics: in the anti-integrable regime $a=0$
the dynamics is spatially decoupled, and instability is maximal since the individual $T_N$ models on an interval yield strong chaos. On the other hand, spatial interplay grows
complex as the coupling $a$ is increased, so that these simplest synchronized states become less unstable, and eventually achieve maximum stiffness in the 
limit $a\rightarrow1$.

 \subsection{Period-2 synchronized state} 
 \label{p2UPO}
 The next case is that of a temporal period-2 state, where, in the language of 
 symbolic dynamics, the potential may take on two values, 
 $V(\phi_{t})=V_{01}$  or $V_{10}$, 
 same for all $n$'s and at alternate times $t$. This periodic orbit is the solution to the system
 \bea 
 \nonumber
 \ssp_{10} &=& 
    (1-a)T_N(\ssp_{01}) 
    + \frac{a}{2}(\ssp_{01} + \ssp_{01}) \\
  \ssp_{01} &=& 
    (1-a)T_N(\ssp_{10}) 
    + \frac{a}{2}(\ssp_{10} + \ssp_{10})     
\,. 
\label{ChebCMLp2SS}
\eea
For the Chebyshev $T_2(\phi)=2\phi^2-1$ model, that reads
\beq
\phi_{01/10} = \frac{-1-a\pm\sqrt{5-18a+9a^2}}{4-4a}
\,.
\ee{p2sol}  
 Importantly, this synchronized state is real valued only in the interval $a\in[0,1/3]$. The general entry~(\ref{Jmn}) of $\tilde{{\jMorb}}$ may thus be written as
 \beq
 \tilde{{\jMorb}}_{m,m'}  = \left[e^{-ik_1'}-1 -a(\cos k_2' -1)\right]\delta_{m',m}  -\frac{1}{2}\left[ \sum_{t\,\mathrm{odd}} e^{i(k_1-k_1')t}V''_{01} + \sum_{t\,\mathrm{even}} e^{i(k_1-k_1')t}V''_{10}\right]\delta_{m_2',m_2}
 \,, 
 \label{p2Jmn}
 \eeq
 since the Fourier transforms over the spatial variable $n$ can be performed by factoring out the sums over $t$, which are split into even and odd contributions. 

Now consider the simplest {\pcell} that can fit this state, that is $[L\times2]$. In this case, $m_2=0,1,...,L-1$, while $m_1=0,1$. Then, the split Fourier transforms in Eq.~(\ref{p2Jmn}) only count one term each, with phase 
\beq
k_1 - k_1' = \left\{
\begin{array}{cl}
0 & m_1-m_1'=0 \\
\pm\pi & m_1-m_1'=\pm 1
\end{array}
\right. 
\,.
\eeq
Then the spatiotemporal {\jacobianOp} in reciprocal space has diagonal entries
\beq
\tilde{{\jMorb}}_{m,m}  = e^{-ik_1}-1 -a(\cos k_2 -1)  -\frac{1}{2}\left[ V''_{01} + V''_{10}\right]
 \,, 
\eeq
and off-diagonal entries 
\beq
\tilde{{\jMorb}}_{m_2\,m_1,m_2\,(m_1\pm1)}  =  -\frac{1}{2}\left[ V''_{01} - V''_{10}\right]
\,.
\eeq
The eigenvalue $\Lambda_{k_1,k_2}$ is evaluated via the relation $\tilde{{\jMorb}}v_{k_1,k_2} = \Lambda_{k_1,k_2}v_{k_1,k_2}$, where $v_{k_1,k_2}$ is a vector of 
$2L$ components (recall that $\tau=2$), which we may take in the form $v_{k_1,k_2} = (a_{+},a_{-},a_{+},a_{-},...,a_{+},a_{-})^T$.  Because the only non-zero entries of the 
{\jacobianOrb} in reciprocal space are $\tilde{{\jMorb}}_{m_2\,m_1,m_2\,m_1}$ and  $\tilde{{\jMorb}}_{m_2\,m_1,m_2\,(m_1\pm1)}$, every row of $\tilde{{\jMorb}}$ 
has but two non-zero entries, and we end up with two coupled equations, one coming from the $(m_2,0)-$th row, and the other from the $(m_2,1)-$th row, so to speak (henceforth drop the index $m_2$ to simplify the notation, so as to keep track of $m_1,m_1'$ here below)
\bea
 \tilde{{\jMorb}}_{0,0}\,a_{+} +  \tilde{{\jMorb}}_{0,1}\,a_{-} &=& \Lambda_{k_2,k_1}\,a_{+} \\ 
 \tilde{{\jMorb}}_{1,1}\,a_{-} +  \tilde{{\jMorb}}_{1,0}\,a_{+} &=& \Lambda_{k_2,k_1}\,a_{-}
 \,.
\eea
That brings a quadratic defining equation for $\Lambda_{k_2,k_1}$, whose solution reads
\beq
\Lambda_{k_2,k_1} = \frac{\tilde{{\jMorb}}_{0,0}+ \tilde{{\jMorb}}_{1,1}\pm\sqrt{(\tilde{{\jMorb}}_{0,0}- \tilde{{\jMorb}}_{1,1})^2+4 \tilde{{\jMorb}}_{0,1} \tilde{{\jMorb}}_{1,0}}}{2}
\,.
\label{p2evals}
\eeq
As done for the steady state, let us now compute the Bravais stability of the period-2 synchronized state. Both Bloch bands computed in~(\ref{p2evals}) contribute:
\beq
 \lambda = \frac{1}{4\pi^2} \int_{-\pi}^\pi dk_2 \int_{-\pi/2}^{\pi/2} dk_1 (\ln |\Lambda_{k}^+| + \ln |\Lambda_{k}^-|) 
 \,.
 \label{p2stab}
 \eeq  
It is noted that the domain of integration of $k_1$ is now reduced to $[-\pi/2,\pi/2]$, since 
the {\jacobianOp} is invariant under translations of time-period  
$\tau=2$. 
Each term in Eq.~(\ref{p2stab}) brings an integral of the form
\beq
 I(a) = \int_{-\pi/2}^{\pi/2} dk_1\ln |g(a,k_2)  \pm \sqrt{4e^{-2ik_1} + \Delta V''^2}|
 \label{2pk1int}
 \,,
 \eeq
 with
 \bea
 g(a,k_2) &=&  -2 -2a(\cos k_2 -1) - (V''_{01} + V''_{10}) \\
 \Delta V'' &=& V''_{01} - V''_{10}
 \eea 
As in the case of the steady state, the integral~(\ref{2pk1int}) is differentiated, and evaluated on (one-half) a closed loop
\beq 
\frac{dI(g)}{dg} = \frac{1}{2\pi} \frac{1}{2}\oint_ {\Gamma}  \frac{dz}{iz} \frac{1}{g(a,k_2) \pm \sqrt{4z^2+  \Delta V''^2}}
 \,, \hspace{1cm}
  \vcenter{\includegraphics[width=0.2\linewidth]{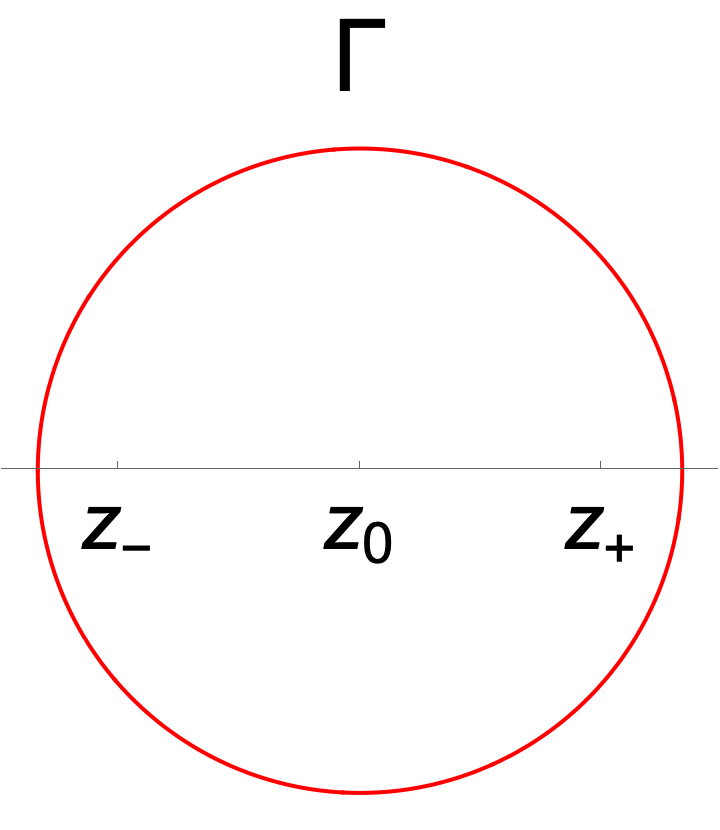}}
\label{dIdg}
\eeq
When rationalized, the integrand is found  to have poles at
\beq
z_0 = 0 \hspace{1cm}
z_{\pm} = \pm \frac{\sqrt{g^2(a,k_2) -   \Delta V''^2}}{2} 
\label{poles}
\,.
\eeq
While $z_0$ is always enclosed by the unit circle in the complex  plane, we need $|z_{\pm}(a,k_2)|<1$ in order for these poles to contribute to the integral.
 \begin{figure}[tbh!]
 \centerline{
(a)\scalebox{.55}{\includegraphics{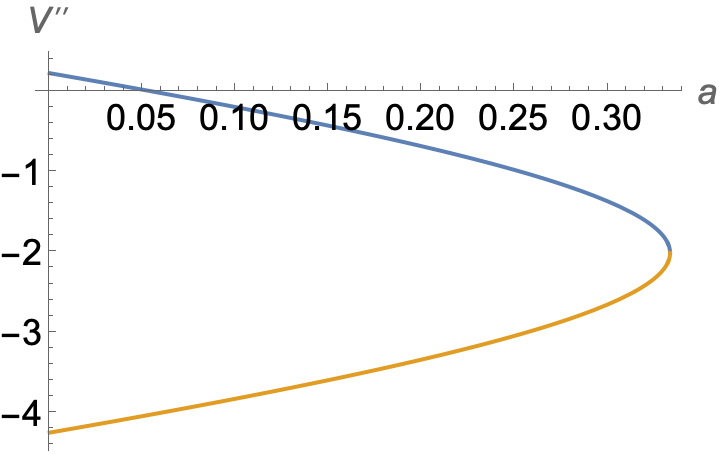}} 
(b)\scalebox{.75}{\includegraphics{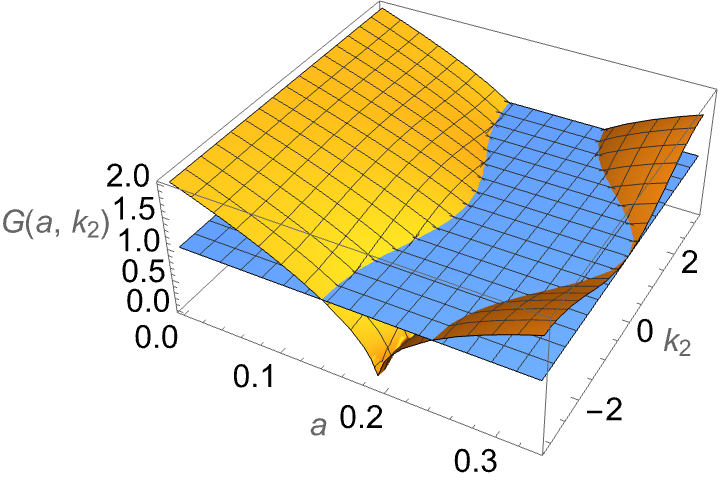}}}
\caption{(a) The expressions $V''_{01}$ (blue) and  $V''_{10}$ (yellow) vs. coupling $a$. (b) The range of $a$ and $k_2$ that make $|z_{\pm}(a,k_2)|<1$ is determined by the intersection of the function $G(a,k_2)=\sqrt{g^2(a,k_2)-\Delta V''}/2$ (in orange, $T_2$ model) with the plane $z=1$ (in blue). When $G(a,k_2)$ lies above the plane,
only one pole contributes to the contour integral~(\ref{dIdg}), otherwise all three poles in Eq.~(\ref{poles}) do, which changes the outcome of the Bravais stability.}
\label{gak2T2}
\end{figure}
From the numerics in~\reffig{gak2T2}, it is noted that for weak lattice couplings only the pole $z_0$ contributes to the integral
and thus to the Bravais stability of the orbit, whereas in the interval $a\in[0.1835,0.2404]$
the contribution comes from all three poles in~(\ref{poles}) \textit{for all} $k_2$. In that interval of the coupling strength, 
the synchronized state~(\ref{p2sol}) is \textit{stable} to perturbations 
of all spatial frequencies $k_2$~(\refref{Dettmann02}). 
As a consequence, we should expect this periodic orbit to also be stable under incoherent perturbations, meaning $\lambda=0$, within the same interval.    

In the light of this observation, the evaluation of the double integral~(\ref{p2stab}) is split into the following intervals:
\begin{enumerate}
\item $a\in[0,0.1409]$: there is only the pole $z_0=0$ within the unit circle, so that
\beq
\frac{dI^\pm(g)}{dg}  =\frac{1}{2} \frac{1}{g(a,k_2) \pm \Delta V''} 
\,,
\label{onepoledIdg}
\eeq
\beq
I^\pm(g) = \frac{1}{2}\ln |g(a,k_2) \pm \Delta V''| + C_\pm =  \frac{1}{2}\ln |(a-1)-V''_\pm -a\cos k_2| 
\,,
\eeq
where $V''_+=V''_{10}$ and $V''_-=V''_{01}$, in the above notation. The additive constant $C_\pm=-\frac{1}{2}\ln2$ is determined so that
\beq
\lambda_+(a=0)+\lambda_-(a=0) = I^+(a=0,k_2) + I^-(a=0,k_2) = \frac{1}{2}\ln|T'(\phi_{01})T'(\phi_{10})|
\,,
\eeq
as one would expect in the zero-coupling ($a=0$) case. Then
\beq
\lambda_\pm = \frac{1}{2\pi} \int_{-\pi}^{\pi} dk_2 \frac{1}{2}\ln |(a-1)-V''_\pm -a\cos k_2| = 
\frac{1}{2}\ln \frac{a-1-V''_\pm + \sqrt{(a-1-V''_\pm)^2-a^2}}{2}
\label{lamSma}
\,,
\eeq  
and $\lambda=\lambda_+ + \lambda_-$, with the outcome is shown in \reffig{lamp2p1T2}(a) in the blue solid line. A  sharp
decrease of the instability with the coupling is noted.
 \begin{figure}[tbh!]
 \centerline{
(a)\scalebox{.55}{\includegraphics{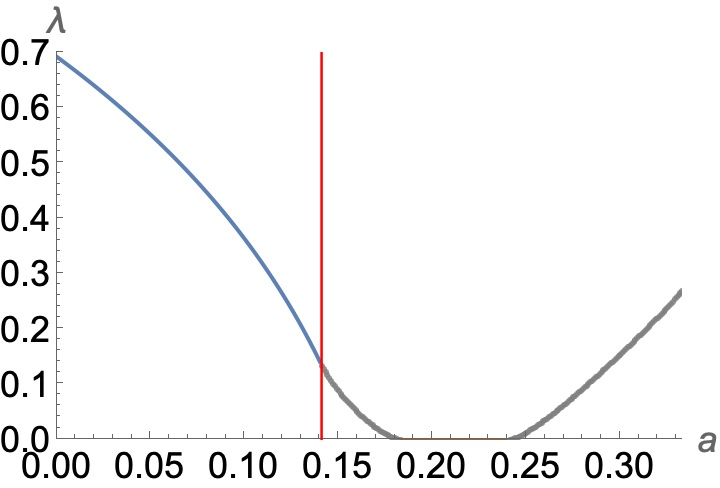}}
(b)\scalebox{.55}{\includegraphics{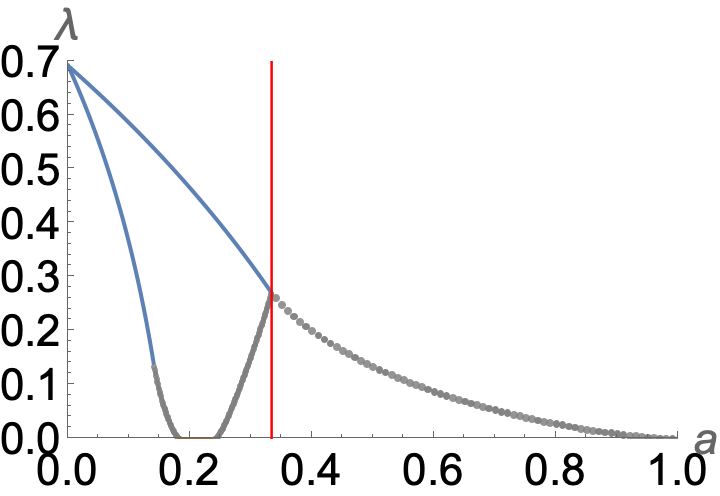}}}
\caption{(a) The stability exponent $\lambda$ versus strength of coupling $a$ for the period-2 synchronized state in the $T_2$ model. (b) Comparison of $\lambda$ with the steady synchronized state $\phi=-1/2$ of the same model.}
\label{lamp2p1T2}
\end{figure}
\item $a\in[0.1409,0.1835]$: the integral in $dk_1$ may have either one or three poles within the unit circle, depending on the value of $k_2$.
There are three poles in the interval $k_2\in]-\pi,-k_2^*]\cup[k_2^*,\pi]$,  where $k_2^*$ solves the equation $\frac{\sqrt{g^2(a,k_2) -   \Delta V''^2}}{2} =1$. 
When $a\simeq0.1409$, $k_2^*\simeq\pi$ and there is still only the pole $z_0=0$ contributing to the integral. As $a$ increases from that value, $k_2^*$ decreases,
and the contributions of the poles $z_\pm$ add up to that of $z_0$ to yield 
\beq
\frac{dI^\pm(g)}{dg}  = \frac{1}{2}\frac{1}{g(a,k_2) \pm \Delta V''} - \frac{1}{2}\left(\frac{g(a,k_2)}{g^2(a,k_2)- \Delta V''^2} \mp \frac{g(a,k_2)}{g^2(a,k_2)- \Delta V''^2} \right)
\,.
\label{dIdg_ctb}
\eeq
When computing $\lambda_+$, the term in parenthesis vanishes, which results in the expression~(\ref{onepoledIdg}), leading to Eq.~(\ref{lamSma})
with a $+$ sign.  The other Bloch band gives $\lambda_-$ by summing up all terms of $\frac{dI^-(g)}{dg}$ in Eq.~(\ref{dIdg_ctb}), and integrating over $g$.   
The result is an integral in $dk_2$ split over the interval $[-\pi,\pi]$, with one pole contributing in $[-k_2^*,k_2^*]$, and three poles in the complement: 
\beq
 \lambda_- = \frac{1}{2\pi} \left[ \int_{-k_2^*}^{k_2^*} \frac{1}{2} \ln |(a-1)-V''_{01} -a\cos k_2|dk_2 -2\int_{k_2^*}^{\pi}\frac{1}{2} \ln |(a-1)-V''_{10} -a\cos k_2|dk_2  \right] 
\,.
\label{k2depint}
\eeq
Combining with $\lambda_+$ [Eq.~(\ref{lamSma})], one obtains
\beq
 \lambda_+  + \lambda_- = \frac{1}{2\pi} \left[ \int_{-k_2^*}^{k_2^*}\frac{1}{2}\ln |(a-1)-V''_{01} - a\cos k_2| dk_2  + \int_{-k_2^*}^{k_2^*}\frac{1}{2} \ln |(a-1)-V''_{10} -a\cos k_2| dk_2  \right] 
\,.
\label{lamMida}
\eeq
The above integrals are evaluated numerically and the outcome is shown again in~\reffig{lamp2p1T2}\,(a) in the dotted gray line that begins at the red vertical line, with the stability further decreasing to vanish at the end of the interval.  
\item $a\in[0.1835,0.2404]$: $k_2^*$ vanishes and therefore the three poles contribute to the contour integrals over the whole $[-\pi,\pi]$ interval. Consequently, 
the $\lambda_+$ and $\lambda_-$ contributions cancel each other and the Bravais stability to incoherent perturbations is identically zero [just set $k_2^*=0$ in the
integrals of Eq.~(\ref{lamMida})]. This result is consistent with the finding\rf{Dettmann02} that this synchronized state becomes stable to perturbations of all 
space frequencies $k_2$ in this domain of coupling strengths.  
\item $a\in[0.2404,1/3]$: the configuration of the poles of the contour integral~(\ref{dIdg}) is complementary  to the one
found for the case $a\in[0.1409,0.1835]$ [Fig.~\ref{gak2T2}\,(b)], therefore the limits of integration of the $\lambda_-$ contribution to the stability exponent are swapped in the two integrals from Eq.~(\ref{k2depint}):
\beq
 \lambda_- = \frac{1}{2\pi} \left[ -\int_{-k_2^*}^{k_2^*}\frac{1}{2} \ln |(a-1)-V''_{10} -a\cos k_2| dk_2  +2\int_{k_2^*}^{\pi}\frac{1}{2} \ln |(a-1)-V''_{01} -a\cos k_2| dk_2  \right] 
\,,
\eeq
Combining with $\lambda_+$, one obtains
\beq
 \lambda_+  + \lambda_- = \frac{1}{2\pi} \left[ \int_{k_2^*}^{\pi} dk_2 \ln |(a-1)-V''_{01} - a\cos k_2| + \int_{k_2^*}^{\pi} dk_2  \ln |(a-1)-V''_{10} -a\cos k_2| \right] 
\,,
\label{lamLa}
\eeq
exhibiting an unexpected increase in instability in this last interval of $a$. 
\end{enumerate}
For comparison, Fig.~\ref{lamp2p1T2}\,(b) features both the behavior of the two Bravais exponents of the steady state $\phi=1$ and of the period-two orbit.
It is noted that the steady state smoothly decreases its instability with the lattice coupling strength, which supports the intuition of collective behavior emerging in
tighter lattices. On the contrary, a synchronized state of time period as low as two is long enough to showcase a nontrivial range of instabilities to both coherent and incoherent    
perturbations. Its Bravais exponent first rapidly drops off to completely vanish at a relatively weak coupling, to then resurge at tighter bindings, before the synchronized state itself
disappears altogether (becoming complex valued) at $a=1/3$.

\section{Discussion}
\label{end}

Nonlinear field theories with defining equations of the type~(\ref{GroBec06:sto2}) feature several interesting phenomena, which can  
efficiently be modeled by means of coupled map lattices\rf{Beck97}.   
In the framework of spatiotemporal chaos, the search for and characterization of recurrent patterns often proves especially insightful\rf{Visw07b,LucKer14,CPTKGS22}.
In particular, understanding the stability of periodic orbits to perturbations is essential for example to separate laminar from turbulent solutions\rf{KK05},
and thus it lies at the core of the dynamical analysis.

In the present work, linear stability analysis has been performed so as to treat space and time on equal grounds. 
This approach is made possible by 
the spatiotemporal {\jacobianOp}, in both phase-space and reciprocal lattice,
where frequencies replace space and time to identify perturbations.
Once the type of stability of periodic orbits to periodic (coherent) perturbations of a certain space or time frequency is determined, 
the stability to aperiodic (incoherent) perturbations is also computed by summing all the eigenvalues of 
the {\jacobianOp} over all frequencies.     

In the specific model of coupled Chebyshev maps, changes of stability may occur as a function of the coupling strength, and, as shown in this venue,
the shortest synchronized states of this lattice model already showcase a number of different scenarios. Synchronized states are the simplest periodic orbits,
which allow for an almost fully analytic treatment of the problem of linear stability to both coherent and incoherent perturbations. 
Here, doing the calculations mostly by hand enables us to unravel the connections between the stability type of steady states or time period-2 orbits to 
perturbations of definite frequencies, and that to general, aperiodic perturbations, thanks to complex analysis (contour integration).
Regarding the (`Bravais') stability to incoherent perturbations, the findings that stem from the present analysis are expected and intuitive for
the synchronized steady states: stronger coupling enhances collective motion and thus reduces instability from the anti-integrable limit. 
Less obvious is the behavior of the time period-2 
synchronized state, which turns from unstable to stable and then back to unstable as the coupling strength is increased, before becoming complex valued and thus
disappearing from the real lattice space. While the previous observations were made in\rf{Dettmann02} following the temporal analysis of perturbations of 
definite space frequencies, here they are confirmed in the full spatiotemporal picture 
(from the computation of 
the {\jacobianOp}), and extended to
incoherent perturbations. 

Based on the present results, it is legitimate to expect a richer landscape of bifurcations when more complex, non-synchronized 
recurrent spatiotemporal patterns are considered. A hint already is given to us by both the phase diagram of the time period-2 synchronized 
state discussed above, and by its interaction with a steady state, portrayed in Fig.~\ref{lamp2p1T2}(b).  
While the quantities of interest and the salient formulae (such as the double integral~(\ref{BrStbExp}) for the Bravais stability) would be
unchanged, more serious computational hurdles would certainly call for numerics in their evaluations, instead of the elegant formal analysis performed here.                

Linear stability analysis of periodic states, and more in general of recurrent patterns, is instrumental for the computation of partition sums in deterministic chaotic 
field theories\rf{lanCvit07,LC21,CL18}. There, the Bravais stability exponent of each orbit constitutes the weight of the corresponding term in the partition functional,
that generates expectation values for all observables of interest in the theory. For example, in chaotic classical field theories such as Kuramoto-Sivashinskii\rf{Christiansen97},
Yang-Mills, or Navier-Stokes\rf{CFTsketch}, global averages such as Lyapunov exponents or correlation functions can be evaluated by means of spatiotemporally periodic patterns.
In Beck's theory\rf{Beck95} of chaotic strings, a number of parameters of the standard model of particle physics were computed by means of averages of interactions and self-energies with a precision of four digits with direct numerical simulations, while temporal periodic orbit theory could demonstratively attain 13 significant figures for some of the same
quantities on a two-site lattice\rf{DetLip05}. Most recently, spatiotemporal chaos shows up in lattice QED\rf{KiTaHa21}, where stochastic quantization is applied to evaluate the anomalous magnetic moment of the electron by numerical averaging\rf{Kitano24}. Recurrent patterns may prove effective and insightful in this class of problems, as well.        

With that perspective, this work is but the beginning of a systematic study of the spatiotemporal stability properties of 
periodic states in a dissipative, or non phase-space volume preserving, class of systems.

\section{Acknowledgements}
The author thanks P. Cvitanovi\'c for 
valuable discussions.
This work has been 
supported by JSPS KAKENHI Grant No. 25K07154  (PI: A. Shudo).

\appendix
\section{Appendix: Evaluation of the Bravais stability exponent for a steady state}
\label{AppA}
 Let us evaluate the integral~(\ref{k2int}) with the residue theorem, dropping the absolute value in the argument of the logarithm: 
 \beq
 \lambda =  \frac{1}{2\pi} \int_{-\pi}^{\pi} dk_2 \ln \left[(1-a)T_N'(\phi_p)+a\cos k_2\right]
 \,.
 \label{k2int2}
 \eeq
 Now 
 call $\xi(a)=(1-a)T_N'(\phi_p)$, and evaluate the quantity
 \beq
 \frac{d\lambda}{d\xi} =   \frac{1}{2\pi} \int_{-\pi}^{\pi}\frac{dk_2}{\xi(a)+a\cos k_2}
 \,.
 \label{dlamdh}
 \eeq
The previous is rewritten as the contour integral (let $z=e^{ik_2}$)
\beq
  \frac{1}{2\pi} \oint_ {\Gamma}  \frac{dz}{iz} \frac{2}{2\xi(a) + a(z + z^{-1})}
  \hspace{1cm}
  \vcenter{\includegraphics[width=0.2\linewidth]{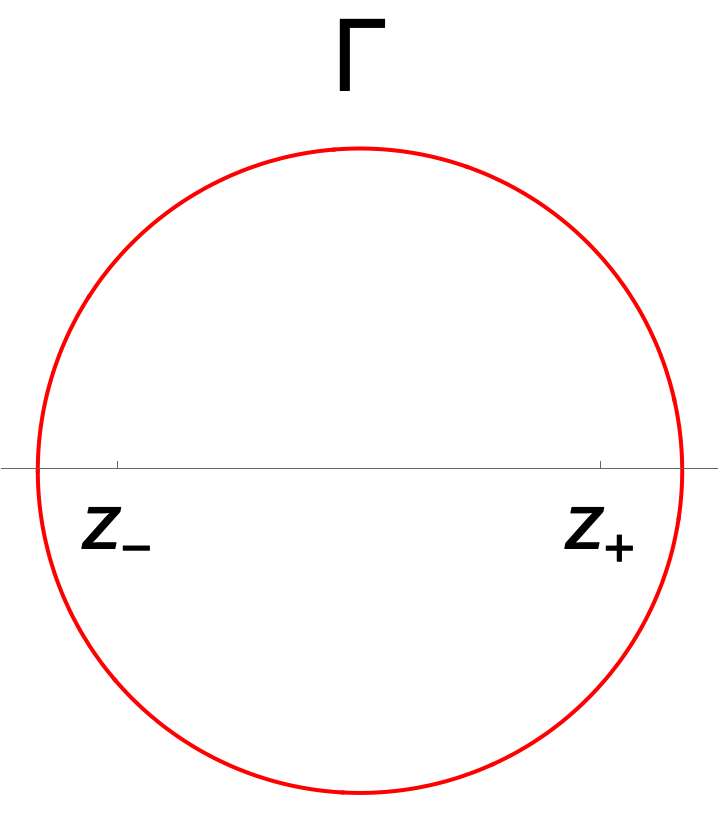}}
  \,,
  \label{circlam}
  \eeq 
The integrand has two poles, $z_{\pm} = \frac{-\xi(a)\pm\sqrt{\xi^2(a)-a^2}}{a}$, both real-valued for $\xi^2(a)-a^2>0$, or, equivalently,
\beq
a <  \frac{|T_N'(\phi_p)|}{1+|T_N'(\phi_p)|}
\,.
\label{realroot}
\eeq
 One can now evaluate~(\ref{circlam}) using the residue theorem:
 \beq
  \frac{d\lambda}{d\xi} =  \frac{1}{2\pi}  \oint \frac{dz}{2\xi(a)z + \frac{a}{2}z^2 + \frac{a}{2}}  = 
   \frac{1}{2\pi}  \oint \frac{dz}{\frac{a}{2}(z-z_-)(z-z_+)}
   \,. 
 \eeq
 How many poles lie within the unit circle? Requiring $|z_\pm|<1$, it turns out that there is only one 
 pole (either $z_+$ or $z_-$) within the unit circle in the interval of $a$ verifying inequality~(\ref{realroot}). 
Then, 
 \beq
  \frac{d\lambda}{d\xi} = \pm\frac{1}{\frac{a}{2}(z_- - z_+)} = \pm \frac{1}{\sqrt{\xi^2(a)-a^2}}
  \,. 
  \label{circlamres}
 \eeq
 Integrating Eq.~(\ref{circlamres}) with respect to $\xi$ and imposing the boundary condition $\lambda(a=0)=\ln|T_N'(\phi_p)|$ 
 (uncoupled lattice), we retrieve the result~(\ref{k2res}):
 \beq
 \nonumber
 \lambda = \ln \frac{(1-a)|T_N'(\phi_p)| + \sqrt{(1-a)^2T_N'^2(\phi_p) - a^2}}{2}
 \,.
 \tag{\ref{k2res}}
 \eeq 
 Importantly, $\lambda$ is a real number in the whole interval of validity of the previous expression, that is for 
 \beq
 a < \frac{|T_N'(\phi_p)| -1}{|T_N'(\phi_p)|+1}
 \,,
 \label{redline}
 \eeq
marked by the red vertical line in Fig.~\ref{lamwholeT2}. The previous condition comes from Eq.~(\ref{k1intcond}) for the integral in the
time-frequency $k_1$ to be nonzero, in which one sets $\cos k_2=\pm1$, depending on the sign of $T'_N(\phi_p)$.  
 
 Beyond that value of $a$, the limits of integration reduce to $-k_2^*, k_2^*$ defined by Eq.~(\ref{k2st}):  
  \beq
 \lambda =  \frac{1}{2\pi} \int_{-k_2^*}^{k_2^*} dk_2 \ln[(1-a)T_N'(\phi_p)+a\cos k_2]
 \,.
 \label{k2stint}
 \eeq
 In the main text, the integral of the $\ln|...|$ was computed numerically, since we are interested in the real part of the stability exponent $\lambda$.
 However, one may wonder if and for what couplings the argument of the log turns negative in Eq.~(\ref{k2stint}), and consequently $\lambda$ acquires an
 imaginary part.
 From the previous expression, we may infer that the argument of the logarithm is positive for
 \beq
 a< \frac{|T_N'(\phi_p)|}{|T_N'(\phi_p)|+1}
 \,,
 \label{logreal}
 \eeq  
 since $\cos k_2\in[\cos k_2^*,1]$, depending on $k_2^*$, so that Eq.~(\ref{logreal}) yields the smallest $a$ that produces a 
 complex-valued $\lambda$. Recalling Eq.~(\ref{redline}), we may conclude  that for
 \beq
 \frac{|T_N'(\phi_p)| -1}{|T_N'(\phi_p)|+1} < a <  \frac{|T_N'(\phi_p)|}{|T_N'(\phi_p)|+1}
 \,,
 \eeq
 the Bravais stability exponent is real valued. The apparent inflection point (Fig.~\ref{lamwholeT2}) observed for values of the
 coupling strength across condition~(\ref{redline}) is then due to the functional change of $\lambda(a)$: from equation~(\ref{k2int2})
 with fixed limits of integration, to Eq.~(\ref{k2stint}), where the $a-$dependence is the same in the integrand, but now it also appears in the 
 the limits of integration, with $k_2^*$ a function of $a$ through Eq.~(\ref{k2st}).

  Let us get as close as possible to a primitive for the integral~(\ref{k2stint}): 
 leverage the trigonometric identity $\cos k_2 = 1 - 2\sin^2 \frac{k_2}{2}$, and operate the substitution
 $P^2 = 4\sin^2 \frac{k_2}{2}$. That way, the integral~(\ref{k2stint}) is rewritten as
 \beq
 \lambda = \frac{\ln a/2}{2\pi} \int_{-P^*}^{P^*}  dP  \frac{\ln(\mu^2 - P^2)}{\sqrt{1-P^2/4}}
 \label{psqint}
 \,, 
 \eeq  
 where $\mu^2 = 2\frac{(1-a)T_N'(\phi_p) + a}{a}$. Splitting $\mu^2-P^2 = (\mu - P)(\mu + P)$,  integrating by parts, and regrouping, produces the outcome
 \beq
 \lambda = \frac{\ln a/2}{2\pi} \left[ \left.\arcsin P \ln (\mu^2-4P^2)\right|_{-P^*/2}^{P^*/2}  + 2 \int_{-P^*/2}^{P^*/2}  dP  \frac{P\arcsin P}{\mu^2-4P^2} \right]
 \,,
 \eeq 
 where the latter term may be written in terms of polylogarithmic functions. Looking at the integrand of Eq.~(\ref{psqint}) and recalling the definition of $\mu$, 
 we once again obtain the condition~(\ref{logreal}) for the logarithm and thus the Bravais exponent to be real valued. For stronger coupling, the stability exponent 
 becomes complex valued, and  thus perturbations drive the field away from the synchronized state $\phi_p$ but this time in an oscillatory fashion.

\printcredits

\bibliographystyle{cas-model2-names}
\bibliography{siminos}

\end{document}